\documentclass[article,twocolumn,showpacs]{revtex4-1}
\usepackage[utf8]{inputenc}
\usepackage{geometry}
\usepackage{amssymb}
\usepackage{graphicx}
\usepackage{amsmath}
\usepackage{float}
\usepackage{times}
\usepackage{color}
\usepackage{subfigure}
\usepackage{setspace}
\usepackage{bm}

\usepackage{epstopdf}
\usepackage{color}
\usepackage{bm}
\usepackage{graphicx}
\usepackage{amsmath}
\usepackage{amsfonts}
\usepackage{amssymb}
\usepackage{bezier} 
\usepackage{color} 

\font\elevenmib=cmmib10 scaled 1095
\font\tenmib=cmmib10
\font\eightmib=cmmib10 scaled 800
\font\sixmib=cmmib10 scaled 667
\skewchar\elevenmib='177
\newfam\mibfam
\def\mib{\fam\mibfam\tenmib}
\textfont\mibfam=\tenmib
\scriptfont\mibfam=\eightmib
\scriptscriptfont\mibfam=\sixmib

\def \be{\begin{equation}}
\def \ee{\end{equation}}
\def \bea{\begin{eqnarray}}
\def \eea{\end{eqnarray}}

\def\nd{{\vphantom{\dagger}}}

\def\bR{{\mib R}}

\def\bz{{\mib z}}

\def\bx{{\mib x}}

\def\frac#1#2{{\textstyle{#1 \over #2}}}

\def\cO{{\cal O}}

\def\ve{\varepsilon}

\def\Im{{\rm Im}}
\def\Re{{\rm Re}}

\mathchardef\Gamma="7100

\def\bk{{\bf k}}

\def\bpi{ \mbox{\boldmath{$\pi$}}}

\def\bpi{ \mbox{\boldmath{$\pi$}}}

\def\Im{{\rm Im}}
\def\ve{{\varepsilon}}
\allowdisplaybreaks

\def\ssr#1{{\sss{\rm #1}}}
\def\sc{\ssr{C}}
\def\st{\ssr{T}}
\def\so{\ssr{O}}
\def\ssr#1{{\scriptstyle\rm #1}}
\def\ssf#1{{\scriptstyle\textsf{#1}}}

\def\sssf#1{{\scriptscriptstyle\textsf{#1}}}
\def\ccmc{\chi_\ssf{CMC}}

\def\ssr#1{{\scriptstyle\rm #1}}
\def\ssf#1{{\scriptstyle\textsf{#1}}}

\def\sssf#1{{\scriptscriptstyle\textsf{#1}}}

\def\csr{\sssf{CSR}}

\begin{document}

\title{Degeneracy-projected polarization formulas for Hall-type conductivities}
 \author{Noga Bashan and Assa Auerbach}
\affiliation{Physics Department, Technion, 32000 Haifa, Israel}
 
\date{\today }
\begin{abstract}
Kubo formulas for Hall,  transverse thermoelectric and thermal Hall conductivities are simplified into
on-shell commutators of degeneracy projected polarizations. The new expressions are computationally economical, and apply to general Hamiltonians without a gap restriction. We show that Hall currents in open boundaries are carried by gapless chiral excitations.
Extrapolation of  finite lattice calculations to the DC-thermodynamic limit is demonstrated for a disordered metal.
\end{abstract}
\maketitle

Electric, thermoelectric, and thermal Hall conductivities, a.k.a. $\sigma_{xy}$, $\alpha_{xy}$, and $\kappa_{xy}$, respectively,  
characterize charge and thermal carriers of condensed matter phases, and identify their topology~\cite{TKNN,Kitaev-calc,Read-Green}.  
Anomalous  Hall and thermal Hall behavior have been reported in strongly interacting systems e.g. cuprates~\cite{Taillefer-Sxy, Taillefer-Kxy} and correlated insulators~\cite{Matsuda-Kitaev,Behnia}.
In principle, they might be explained by computing Kubo formulas~\cite{Kubo,Lutinger,Cooper}.

Unfortunately, DC Hall-type Kubo formulas are computationally costly.  Their {\em off-shell}  (energy non-conserving) matrix elements of the currents, require full diagonalization of the Hamiltonian on large systems.  In addition, divergent magnetization subtractions~\cite{Cooper} for thermal Hall coefficients require careful cancellation~\cite{Niu,Cong}.  

Berry curvature (Chern) integrals \cite{TKNN,yosi} and  Streda (equilibrium) formulas \cite{Streda,Streda-Kxy},  approximate the Kubo formula by
{\em reversing} the DC order of limits (i.e. setting frequency to zero  before taking the large volume limit~\cite{EMT}). Thus, they only apply to bulk-gapped phases with vanishing longitudinal conductivities, e.g. Quantum Hall (QH) and topological insulators (TI).

This paper  simplifies the Kubo formulas in the proper DC order of limits. 
The new formulas are compact and valid for general Hamiltonians, including gapless phases with
disorder and interactions. Physical insight is gained by  
expressing $\sigma_{xy}$, $\alpha_{xy}$ and $\kappa_{xy}$ as commutators of
degeneracy-projected polarizations (DPPs). The DPPs generalize the role of  Landau guiding centers to gapless phases. 
The formulas imply that Hall and thermal Hall currents  are carried by extended chiral excitations which
may be supported on the sample edges (for e.g. QH and TI) or may percolate through the bulk.

The conductivities are expressed by a smaller sum over {\em on shell} matrix elements, which is computationally  economical.
Problematic magnetization subtractions in the thermal conductivities are eliminated. 
At low temperatures, the relevant eigenstates are confined  to low energies, which allows one to replace the microscopic model by 
a simpler low energy effective Hamiltonian.

Finite lattice calculations  require extrapolation to the DC-thermodynamic limit.
A finite size scaling scheme is demonstrated  for the metallic phase of disordered electrons at weak magnetic fields. The numerical results
recover Drude-Boltzmann (DB) theory, and  Wiedemann-Franz  law  for that model.

{\em Kubo formulas} -- 
The  DC-thermodynamic limit of transport coefficients is defined by,
\be
S^{xy-{\rm dc}}_{\so\so'} \equiv \lim_{\stackrel{ \ve\to 0}{V\to \infty}} S^{xy}_{\so\so'}( \ve ,  V),~~~\so,\so'=\sc,\st,
\label{DC}\ee
where the charge (C) and thermal (T)  Hall-type conductivities are 
 $\sigma_{xy}\equiv S^{xy}_{\sc\sc}$, $ \alpha_{xy}\equiv S^{xy}_{\sc\st}/T$, and $ \kappa_{xy}\equiv S^{xy}_{\st\st}/ T$.
$ \ve,V$ are finite imaginary frequency and volume respectively. The DC order of limits is taken from bottom to top~\cite{Comm-OOL}.

Here we consider a general many-body lattice Hamiltonian  $H$ on open boundary conditions  (OBC) \cite{Comm-OBC},  
C4 symmetry in the $xy$  plane \cite{Comm-C4}, and a magnetic field $B$ in the $z$ direction.
Its spectrum and eigenstates are $\{E_n,|n\rangle\}$. 
The   Hall-type Kubo formulas in the Lehmann representation are,
\be
S^{xy}_{\so\so'}= {\hbar \over V}   \Im  \sum_{n, m} \! { (\rho_m\!-\!\rho_n )  
 \langle m | {j^x_{\so}}|n\rangle\langle n|  j^y_{\so'}  |m\rangle \over  (E_n\!-\!E_m)(E_n\!-\!E_m   -i\ve)  } 
-  { \langle M_{\so\so'}\rangle \over V}.
\label{Kubo}
\ee
$\rho_n(T)$ are Boltzmann weights at temperature $T$.
The magnetization terms $\propto{ \langle M_{\so\so'}\rangle }$  eliminate circulating magnetization currents from the first term~\cite{Cooper}.

The currents $j^\alpha_\so$ and magnetizations $M_{\so\so'}$ are defined as follows.
The Hamiltonian is spatially decomposed on the lattice $H = \sum_i h_i$~\cite{Comm-SD}.
The charge and thermal polarizations are,
\be 
 P_\sc^\alpha \equiv e  \sum_{i} n_i x_i^\alpha ,\quad P^\alpha_\st=  \sum_{i} h_i x_i^\alpha,\quad \alpha=x,y,
\ee
where $e n_i$ is the local charge density, and $\bx_i$ is the position  of lattice site  $i$. 
The electric and thermal currents  are,
\be
 j_\so^\alpha = {i\over \hbar}  [H, P_\so^\alpha], \quad  \so=\sc,\st. \label{qeq0}
\ee
 In the literature one often finds first quantized expressions for the magnetizations
~\cite{Comm-Particles}. Here we use more general definitions which apply to any form of the hamiltonian,
\be
M_{\sc\st} = -{i \over \hbar}[P_\sc^x, P_\st^y],\quad 
M_{\st\st} = -  { i \over \hbar }[P_\st^x, P_\st^y] .
\label{Mqeq0}
\ee
Note that $M_{\sc\sc}=0$, since the two charge polarizations commute.
For anomalous bosonic Hamiltonians $\langle M_{\st\st}\rangle/T$  may diverge  as $\lim_{T\to 0} $. Such divergence must be precisely cancelled
by the current correlators, as shown for non-interacting QH systems~\cite{Read-Kubo,Niu}. Such cancellations could be problematic if one  applies
separate approximations to the two terms in Eqs.~(\ref{Kubo}).

{\em DPP formulas} -- Eqs.~(\ref{Kubo}) are simplified as follows. 
The real part of the summands'  numerator  vanishes  by C4 symmetry,
\be
 \Re  \langle m|j_\so^x|n\rangle\langle n|j_{\so'}^y|m\rangle=0.
\ee
The real part of the denominator  is written as two terms,
\be 
\Re {1 \over \Delta_{nm}(\Delta_{nm}-i\ve) } = {1 \over \Delta_{nm}^2}  - { \ve^2 \over \Delta_{nm}^2 (\Delta_{nm}^2 +\ve^2) } ,\label{trick}
\ee 
where  $\Delta_{nm}\equiv E_n\!-\!E_m$. The matrix elements of Eq.~(\ref{qeq0}) in the eigenstates basis are,
\be
{\langle n|j_\so^\alpha|m\rangle\over \Delta_{nm}} =  { i \over \hbar} \langle n|P_{\so}^\alpha|m\rangle,
\ee
which we insert into Eq.~(\ref{Kubo}) to yield,
\bea
S^{xy}_{\so\so'}&=&  {1 \over\hbar V}    \sum_{n  m} \!   (\rho_n\!-\!\rho_m )~\Im \Bigg(   \langle m|P_{\so}^x|n\rangle  \langle n|P_{\so'}^y|m\rangle \nonumber\\
 &&    - { \ve^2   \langle m|P_{\so}^x|n\rangle  \langle n|P_{\so'}^y|m\rangle \over (\Delta_{nm}^2 +\ve^2) } \Bigg)-  { \langle M_{\so\so'}\rangle \over V}.
\label{Kubo-0}
\eea
The top row, which is an off-shell sum, can be rewritten as the thermodynamic average of  the polarizations' commutator $\langle [P_\so^\alpha,P_{\so'}^\beta]\rangle$.
Therefore  it vanishes for $\sigma_{xy}$, and 
precisely cancels with the magnetization corrections  (\ref{Mqeq0}) for $\alpha_{xy}$  and $\kappa_{xy}$ (good riddance!).  

{\em Surprisingly,  it is the  seemingly negligible $ \ve^2 $-term which fully determines $S^{xy-{\rm dc}}$~!}
The Kubo formulas  reduce to a purely on-shell expression,
\be 
 S^{xy-{\rm dc}}_{\so\so'}   = - \lim_{ \stackrel{\ve\to 0}{ V\to \infty}}  { 1 \over \hbar V} \Im \sum_{n } \!   \rho_n  \langle n| \left[ \tilde{P}_\so^x ,\tilde{P}^y_{\so'} \right] |n\rangle,
  \label{Kubo1}
\ee 
where $\tilde{P}_\so^\alpha$  is the DPP in the $\alpha$ direction,
\be 
\langle n| \tilde{P}_\so^\alpha |m\rangle  =   \langle n | P_\so^\alpha |m\rangle \Theta_\ve(|E_n-E_m|),
\label{DPP-def}
\ee
and the Lorentzian $\Theta_\ve(x) ={\ve^2\over x^2+\ve^2}$  can be replaced by a projector  Heaviside function $\Theta_\ve(x) \to \Theta(\pi \ve/2-|x|)$ in the limit $\ve\to 0$.

{\em Reduction to single particle (SP) Hamiltonians} -- For non-interacting fermions or bosons,
\be
H^{\rm sp}= \sum_{ij} h_{ij}(B) a^\dagger_i a_j=\sum_{\alpha} \epsilon_\alpha(B) a^\dagger_\alpha a_\alpha,
\ee
Eq.~(\ref{Kubo1}) reduces to
\be
S^{xy-{\rm sp}}_{\so\so'} = -\lim_{ \stackrel{\ve\to 0}{ V\to \infty}} {1\over \hbar V}   \Im \sum_{\alpha} n_\alpha ~\left[ \tilde{P}_{\so}^x ,\tilde{P}^y_{\so'} \right]_{\alpha\alpha} ,
  \label{Kubo1-sp}
\ee
where  $n_\alpha$ is the Fermi-Dirac  or Bose-Einstein  occupation of SP state $a^\dagger_\alpha|0\rangle$. The DPPs are \cite{Comm-anomalous},
\bea
\tilde{P}_{\so}^\gamma &=&  \sum_{\alpha\beta}  \left(\tilde{P}_{\so}^\gamma\right)_{\alpha\beta} a^\dagger_\alpha a^\nd_\beta , \nonumber\\
 \left(\tilde{P}_{\so}^\gamma\right)_{\alpha\beta} &= &  \langle\alpha| P_{\so}^\gamma |\beta\rangle \Theta_\ve(|\epsilon_\alpha-\epsilon_\beta|)  .
\label{DPP-sp}
  \eea
A version of Eq.~(\ref{Kubo1-sp}) was derived by Bradlyn and Read~\cite{Read-Kubo}
for  integer   QH states without disorder.

{\em DPP's in  clean Landau levels} -- Eigenstates of  electrons  of effective mass $m$ in a strong magnetic field 
are described  by degenerate Landau levels.
The  charge polarizations (whose components commute) can be decomposed as  
\be
 P_\sc^\gamma=   e  R^\gamma + e l_B ( \bpi \times \hat{\bz} )^\gamma,
 \label{Pgamma}
\ee
where $l_B=\sqrt{  \hbar c \over e B}$. $\bpi$ connect  between adjacent  Landau levels.  $\bR$ are guiding center coordinates which satisfy $[R^x,R^y]= - i  l_B^2\mathbb{I} $, and $[R^\alpha,\pi^\beta]=0$.

On OBC, $H^{\rm sp}$ includes a confining potential $V(\bx)$ on its edges.
A smooth potential~\cite{Comm-smooth} with $|\nabla\log V|^{-1}\ll l_B$ 
can be approximated by an intra-Landau level operator $V^{\rm eff}(\bR)$.
$V^{\rm eff}(\bR)$ (which commutes  $\bpi$) acts only within a single Landau level labelled by $\nu$. One can choose the eigenstate basis of say $R^y|\nu,k\rangle = k l_B |\nu,k\rangle$,  in which $\langle \nu, k | V^{\rm eff} |\nu, k'\rangle$ is generally not diagonal. $U_{\alpha,k}(V^{\rm eff})$ is the unitary matrix which diagonalizes $V^{\rm eff}(\bR)$, and  defines the energy eigenbasis $|\nu ,\alpha\rangle$. Since  $[R^x,R^y]  = -il_B^2 \mathbb{I}$, we know that  $[U^\dagger R^x U,U^\dagger R^y U] = -il_B^2 \mathbb{I}$.   
The expectation value of the second commutator  is used in
Eq.~(\ref{Kubo1-sp}) to obtain $ \sigma_{xy} =  {nec \over B}$, where $n$ is the electron density. 
This result also holds  in the presence of translationally invariant many body interactions.

{\em   Disordered metals in weak magnetic fields.}--  This regime can be described by DB theory~\cite{Ziman} at small Hall angles $\omega_c\tau \ll 1$,  where $\tau$ is the transport scattering time
and $\omega_c={eB\over mc}$ is the cyclotron frequency.
The DB Hall conductivity yields,
\be
\sigma^{\rm DB}_{xy} = { n e^2 \omega_c \tau^2\over m}.
\label{Drude}
\ee
In the weak field regime, disorder strongly mixes the Landau levels  and severs the relation between the DPPs and the guiding centers. 
 Eq.~(\ref{Drude}) can be recovered by a multiplicative renormalization of the DPPs, i.e. $\tilde{P}_\sc^\gamma \simeq  e ( \omega_c \tau) R^\gamma$, in Eq.~(\ref{Kubo1-sp}).

\begin{figure}[h]
\begin{center}
\includegraphics[width=7cm]{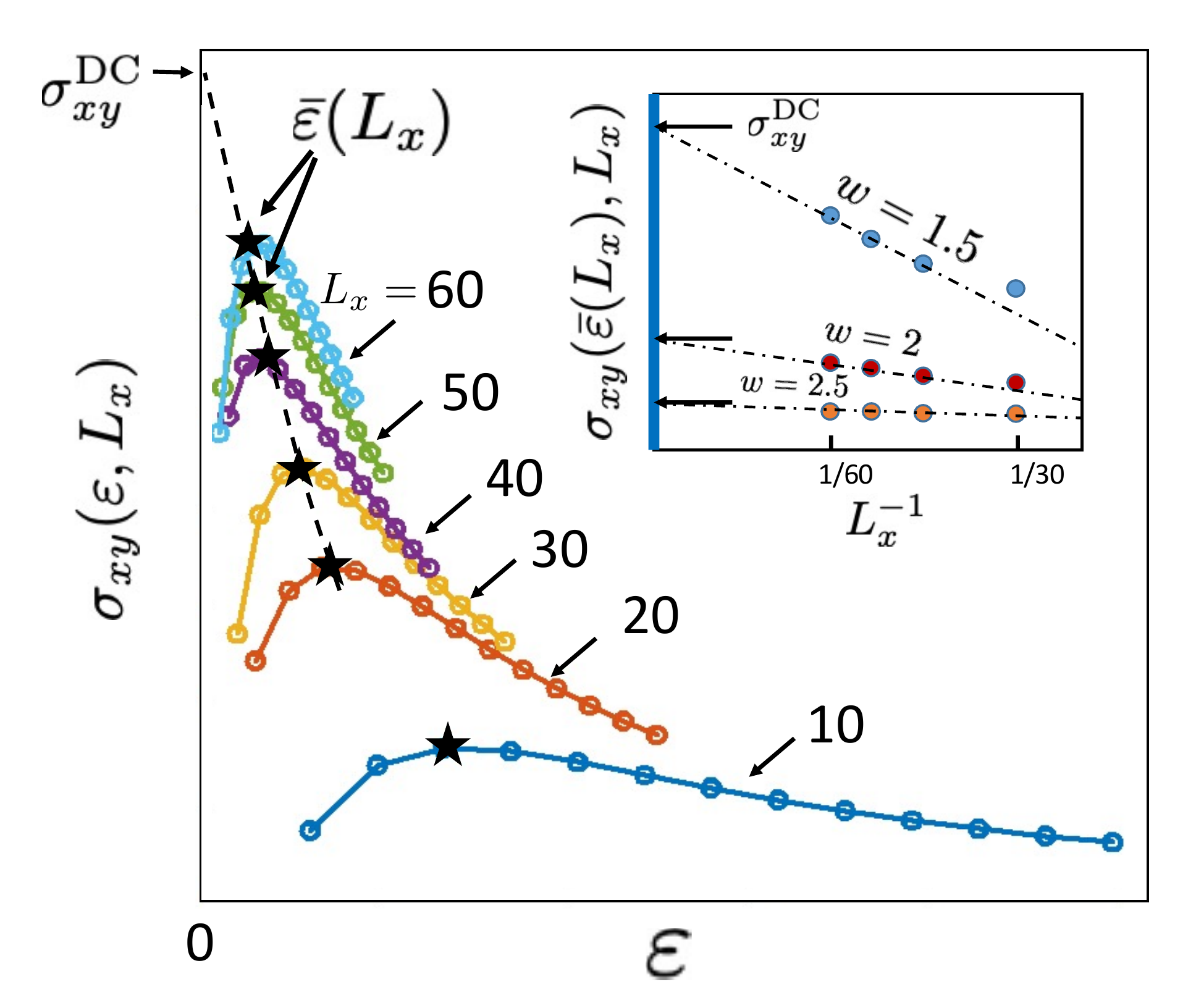}
\caption{Extrapolation of numerical Hall conductivities of the square lattice Hamiltonian, Eq.~(\ref{TB}).
Disorder averaged $\sigma_{xy}$ are plotted versus $\ve$, for a sequence of
linear dimensions $L_x$.  Stars mark the values of $\bar{\ve}$ as defined in Eq.~(\ref{scaling}).  The disorder strength is fixed at $w=3$. 
The temperature, Fermi energy, and magnetic field are $T=0.3,\epsilon_F=-1$ and $B=0.025$
respectively.
{\bf Inset:} The DC limit  $\sigma_{xy}^{\rm dc}$ (marked by black arrows)
for three values of disorder strength. }
\label{fig:scaling}
\end{center}
\end{figure}
{\em Numerical calculations} -- 
Eq.~(\ref{Kubo1}) are significantly less costly than the off-shell formulas Eq.~(\ref{Kubo}).  Having eliminated $-{M_{\sc\st}\over T}, -{M_{\st\st}\over T}$ in $\alpha_{xy}$ and  $\kappa_{xy}$ respectively,
one may apply controlled approximations without worrying about  precise cancellations of divergent corrections.
While Eq.~(\ref{Kubo}) requires full diagnalization  of $H$
 and calculations of  many current matrix elements,  Eq.~(\ref{Kubo1}) includes only matrix elements between nearly degenerate eigenstates
in the spectrum below  temperature $T$.  These states may be numerically accessible by Lanczos algorithms~\cite{Lauchli}  or  approximated by variational methods~\cite{DMRG}. 

Contrary to the initial off-shell formulation of Eq.~(\ref{Kubo}), $H$  in (\ref{Kubo1}) may be replaced 
by its low energy effective Hamiltonian in the spectrum range $E_n-E_0\le k_B T$.  For example,  Fermi liquid theory can be used for interacting fermions,  or continuum field theories for 
magnets and superconductors.

Numerical calculations are mostly performed on finite lattices  which require extrapolation to the DC-thermodynamic limit, of Eq.~(\ref{DC}). 
If $S_{\so\so'}(\ve,L_x )$ is computed
for a sequence of  linear dimensions $\{ L_x^i \}$,  ``optimal'' values of $\bar{\ve}(L^i_x)$ can be extracted by the extrema conditions,
\be
\partial _\ve S^{xy}_{\so\so'}(\ve,L^i_x)=0 \Rightarrow  \bar{\ve}(L^i_x).
\label{scaling}
\ee
The DC limit is obtained by extrapolating  the extrema,  $S^{\rm dc}\!=\! \lim_{i\to \infty} S_{\so\so'}(\bar{\ve}(L^i_x),L^i_x )$.  This scheme is demonstrated
in Fig.~\ref{fig:scaling} 
for the square lattice Hamiltonian,
\be
H^{\rm SL}=-\sum_{\langle ij \rangle}  \left( e^{-iA_{ij}} c^\dagger_i c^\nd_j + {\rm h.c.} \right)+\sum_i ( w_i -\epsilon_F)c^\dagger_i c_i,
\label{TB}
\ee
where $w_i\in[-w/2,w/2]$ is a uniformly distributed random number and $B= \sum_{\square} A_{ij} $ is the magnetic field.

$\bar{\ve}$ are marked by black stars which apparently can be fit to a power law  $\bar{\ve} \simeq 7 (L_x)^{-1}$.
This scaling is consistent with level  spacings  of 
one dimensional extended states. 
In the inset of Figure \ref{fig:scaling}, the Hall conductivities at different disorder strengths  
extrapolate  linearly in $L_x^{-1}$, to their respective DC limits.

In Ref.~~\cite{SM},  we show that $\sigma_{xy}^{\rm dc} \sim w^{-4}$, which is consistent with   DB  result ~(\ref{Drude}),
since by Fermi's golden rule ${1\over \tau}\propto w^2$.  The zero field Hall coefficient $R_{\rm H} = {d\over dB}\sigma_{xy}\sigma_{xx}^{-2}$ varies weakly 
with  $w$ in the moderate disorder regime, as expected by DB theory in the constant life-time approximation.
$R_{\rm H}$ is approximated fairly well by the equilibrium Hall coefficient formula derived in Refs.~\cite{EMT,Abhisek}. In addition, Wiedemann-Franz law 
${\kappa_{xy}\over T\sigma_{xy} } $  reaches close proximity to the DB result of $ { \pi^2  \over 3} $ at low temperatures. 

{\em Discussion} -- 
Since Eq.~(\ref{Kubo1}) applies to any Hamiltonian  with OBC,
we can draw  general conclusions concerning Hall effects  even in strong disorder and interactions regimes:  
(i) Quasi-degenerate manifolds of eigenstates are created by the magnetic field.
 (ii) These manifolds are subjected to {\em non commutative geometry} by the DPPs, i.e. $\langle \left[ \tilde{P}_\so^\alpha ,\tilde{P}^\beta_{\so'} \right] \rangle =  i \epsilon_{\alpha\beta} c $. In a sense,   $ c^{-1} \epsilon_{\alpha\beta}\tilde{P}_\so^\beta=\Pi_\so^\alpha$ generates translations  and acts as a conjugate momentum to $\tilde{P}_{\so'}^\alpha$. Thus, the DPP's  generalize  the algebra of guiding centers to regimes of strong Landau level mixing.
(iii) The quasi-degenerate eigenstates are {\em chiral} as defined by their nonzero  vorticity  as $  \langle\nabla_{\tilde{P}_\so}  \times\vec{ \Pi}_{\so'}  \rangle = 2 /c\ne 0$.
  
These gapless chiral wavefunctions may be supported exclusively on the sample edges, as in QH and TI phases, or in the bulk. 
Bulk chiral states have been derived  semiclassically  by Chalker and Coddington~\cite{CC},
who described the  transition between incompressible plateaux using a percolating network of extended  chiral states.
Finally, we can also infer that thermal Hall currents in insulators~\cite{Matsuda-Kitaev,Behnia,Taillefer-Kxy}  are also carried by extended chiral modes.

 {\em Summary --}  Microscopic computations of charge and thermal Hall conductivities are made easier  by Eqs.~(\ref{Kubo1}, \ref{Kubo1-sp}), which is especially
needed in gapless phases. The new formulas  unveil the essential  role of non-commuting DPPs and associated quasi-degenerate chiral eigenstates.  
 We expect these formulas to allow better connection between model Hamiltonians and
 Hall-type measurements in regimes of strong  scattering.
 
{\em  Acknowledgements} -- We thank D. Arovas,  J. Avron, G. Murthy,  A. Samanta and E. Shimshoni for discussions. 
We acknowledge support from the US-Israel
Binational Science Foundation Grant No. 2016168,  and the
Israel Science Foundation Grant No. 2021367.
This work was performed in part at the Aspen Center for Physics,  supported by National Science Foundation grant PHY-1607611,
and at Kavli Institute for Theoretical Physics, supported  by Grant No. NSF PHY-1748958.

 \appendix
 
\section{Supplementary material}
{\em We demonstrate a numerical calculation of the DC Hall conductivity using  Eq.~(13) of the main text. 
The moderate disorder dependence of $\sigma_{xy}^{\rm DC}(w)$ is shown to scale as $ Bw^{-4}$ as expected by Drude-Boltzmann theory in the
metallic regime. The Hall coefficient is compared to the equilibrium Hall coefficient formula of Ref.~\cite{EMT} at leading order in $w$.
}\\

In the main text we consider the disordered square lattice Hamiltonian, on a lattice of size $L_x^2$,
\be
H^{\rm SL}=-\sum_{\langle ij \rangle}  \left( e^{-iA_{ij}} c^\dagger_i c^\nd_j + {\rm h.c.} \right)+\sum_i ( w_i -\epsilon_F)c^\dagger_i c_i,
\label{TB}
\ee
where $w_i\in[-w/2,w/2]$ is a uniformly distributed random number and $B= \sum_{\square} A_{ij} $ is the magnetic field.

In Drude-Boltzmann theory, the scattering rate is obtained by Fermi golden rule
\be
{\hbar \over \tau} = \pi N(\epsilon_F) w^2
\ee
\begin{figure}[h]
\begin{center}
\includegraphics[width=7cm]{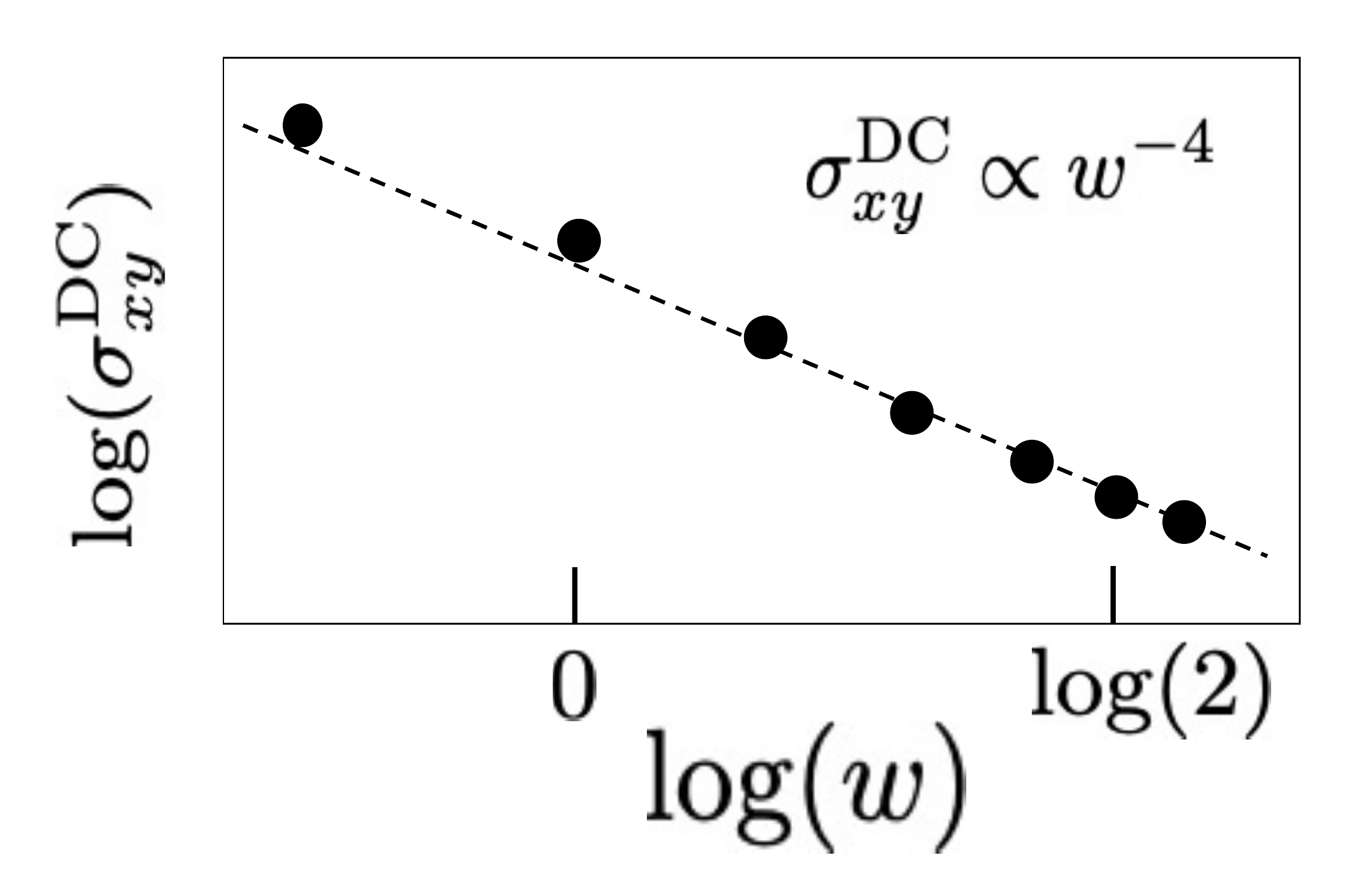}
\caption{ DC extrapolated Hall conductivity for weak magnetic field $B=0.03$
and moderate disorder strength $w$, at Fermi energy $\epsilon_F=-1$. The Hall conductivity scales as $w^{-4}$ as expected by Drude's theory $\sigma_{xy}\propto \omega_c \tau^2$.}
\label{fig:SXY-w}
\end{center}
\end{figure}

Fig.~\ref{fig:SXY-w} shows that the disorder averaged Hall conductivity scales as $w^{-4}$. It is also found to be linear in $B$, which agrees with Drude's theory $\sigma_{xy}\propto \omega_c \tau^2$.

For an additional test,  we also compute the zero field Hall coefficient,
\be
R_{\rm H} (w) = {d\sigma_{xy}\over dB}\sigma_{xx}^{-2}\Big|_{B = 0},
\label{RH-def}
\ee

\begin{figure}[h]
\begin{center}
\includegraphics[width=7cm]{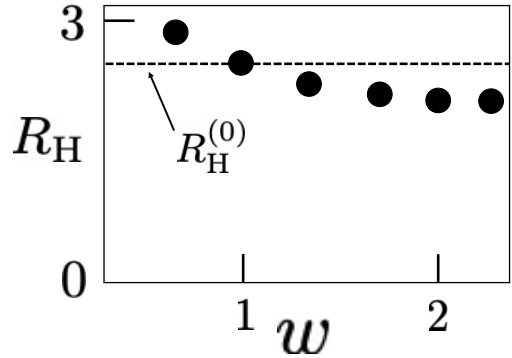}
\caption{ Hall coefficient $R_{\rm H}$ (black circles) of the square lattice model (\ref{TB}) computed by Eq.~(\ref{RH-def}), where $\sigma^{\rm DC}_{xy}$ and $\sigma^{\rm DC}_{xx}$
are extrapolated to $L_x\to \infty$ and disorder averaged.  The   leading order equilibrium formula, Eqs.~(\ref{formula},\ref{sus}), is shown  (dashed line) for comparison.  }
\label{fig:RH-w}
\end{center}
\end{figure}

The equilibrium Hall coefficient formula \cite{EMT,Abhisek} at low disorder is,
\be
R_{\rm H} = {\ccmc\over \chi_\csr^2}~+ \cO(w^2)
\label{formula}
\ee
where the two clean susceptibilities are,
\bea
\ccmc &=&  2\int {d^2 k\over (2\pi)^2}\left(-{\partial f\over \partial\epsilon}\right) \left({\partial \epsilon_\bk\over \partial k_x }\right)^2 
\left({\partial^2 \epsilon_\bk\over \partial k_y^2 }\right), \nonumber\\
\chi_\csr&=&  \int {d^2 k\over (2\pi)^2} \left(-{\partial f\over \partial\epsilon}\right)  \left( {\partial \epsilon_\bk\over \partial k_x }\right)^2,
\label{sus}
\eea
where $f$ is the fermi function at temperature $T$ and Fermi energy $\epsilon_F$. $\epsilon_\bk$ is the square lattice bandstructure,
\be
\epsilon_k = -2 \cos(k_x)-2 \cos(k_y).
\ee
Eqs.~(\ref{formula},\ref{sus}) agree  with DB theory in the constant lifetime approximation \cite{Abhisek}.

In  Fig.~\ref{fig:RH-w}  we plot the disorder averaged  Hall coefficient (\ref{RH-def}).
We note the weak disorder dependence, which is in qualitative agreement with Eq.~(\ref{formula}) and  with DB theory~\cite{Ziman}.

 \bibliographystyle{unsrt}
\bibliography{refs.bib}

\begin{thebibliography}{10}

\bibitem{TKNN}
David~J Thouless, Mahito Kohmoto, M~Peter Nightingale, and Marcel den Nijs.
\newblock {Quantized Hall conductance in a two-dimensional periodic potential}.
\newblock {\em Physical review letters}, 49(6):405, 1982.

\bibitem{Kitaev-calc}
Joji Nasu, Junki Yoshitake, and Yukitoshi Motome.
\newblock {Thermal transport in the Kitaev model}.
\newblock {\em Physical review letters}, 119(12):127204, 2017.

\bibitem{Read-Green}
N.~Read and Dmitry Green.
\newblock Paired states of fermions in two dimensions with breaking of parity
  and time-reversal symmetries and the fractional quantum hall effect.
\newblock {\em Phys. Rev. B}, 61:10267--10297, Apr 2000.

\bibitem{Taillefer-Sxy}
S.~Badoux, W.~Tabis, F.~Lalibert{\'e}, G.~Grissonnanche, B.~Vignolle,
  D.~Vignolles, J.~B{\'e}ard, D.~A. Bonn, W.~N. Hardy, R.~Liang,
  N.~Doiron-Leyraud, Louis Taillefer, and Cyril Proust.
\newblock {Change of carrier density at the pseudogap critical point of a
  cuprate superconductor}.
\newblock {\em Nature}, 531:210 EP --, 02 2016.

\bibitem{Taillefer-Kxy}
G.~Grissonnanche, S.~Th{\'e}riault, A.~Gourgout, M.~E. Boulanger, E.~Lefran{\c
  c}ois, A.~Ataei, F.~Lalibert{\'e}, M.~Dion, J.~S. Zhou, S.~Pyon, T.~Takayama,
  H.~Takagi, N.~Doiron-Leyraud, and L.~Taillefer.
\newblock Chiral phonons in the pseudogap phase of cuprates.
\newblock {\em Nature Physics}, 16(11):1108--1111, 2020.

\bibitem{Matsuda-Kitaev}
Y.~Kasahara, K.~Sugii, T.~Ohnishi, M.~Shimozawa, M.~Yamashita, N.~Kurita,
  H.~Tanaka, J.~Nasu, Y.~Motome, T.~Shibauchi, and Y.~Matsuda.
\newblock Unusual thermal hall effect in a kitaev spin liquid candidate
  $\ensuremath{\alpha}\text{\ensuremath{-}}{\mathrm{rucl}}_{3}$.
\newblock {\em Phys. Rev. Lett.}, 120:217205, May 2018.

\bibitem{Behnia}
Xiaokang Li, Beno{\^\i}t Fauqu{\'e}, Zengwei Zhu, and Kamran Behnia.
\newblock {Phonon thermal Hall effect in strontium titanate}.
\newblock {\em Physical review letters}, 124(10):105901, 2020.

\bibitem{Kubo}
Ryogo Kubo.
\newblock {Statistical-Mechanical Theory of Irreversible Processes. I. General
  Theory and Simple Applications to Magnetic and Conduction Problems}.
\newblock {\em Journal of the Physical Society of Japan}, 12(6):570--586, 1957.

\bibitem{Lutinger}
JM~Luttinger.
\newblock {Theory of thermal transport coefficients}.
\newblock {\em Physical Review}, 135(6A):A1505, 1964.

\bibitem{Cooper}
NR~Cooper, BI~Halperin, and IM~Ruzin.
\newblock {Thermoelectric response of an interacting two-dimensional electron
  gas in a quantizing magnetic field}.
\newblock {\em Physical Review B}, 55(4):2344, 1997.

\bibitem{Niu}
Tao Qin, Qian Niu, and Junren Shi.
\newblock {Energy magnetization and the thermal Hall effect}.
\newblock {\em Physical review letters}, 107(23):236601, 2011.

\bibitem{Cong}
Cong Xiao and Qian Niu.
\newblock Unified bulk semiclassical theory for intrinsic thermal transport and
  magnetization currents.
\newblock {\em Phys. Rev. B}, 101:235430, Jun 2020.

\bibitem{yosi}
J.E. Avron and R.~Seiler.
\newblock {Quantization of the Hall conductance for general, multiparticle
  Schr{\"o}dinger Hamiltonians}.
\newblock {\em Physical review letters}, 54(4):259, 1985.

\bibitem{Streda}
P~Streda and L~Smrcka.
\newblock {Thermodynamic derivation of the Hall current and the thermopower in
  quantising magnetic field}.
\newblock {\em Journal of Physics C: Solid State Physics}, 16(24):L895, 1983.

\bibitem{Streda-Kxy}
Herbert Oji and P~Streda.
\newblock {Theory of electronic thermal transport: Magnetoquantum corrections
  to the thermal transport coefficients}.
\newblock {\em Physical Review B}, 31(11):7291, 1985.

\bibitem{EMT}
Assa Auerbach.
\newblock {Equilibrium formulae for transverse magnetotransport of strongly
  correlated metals}.
\newblock {\em Physical Review B}, 99(11):115115, 2019.

\bibitem{Comm-OOL}
{On OBC, the wavevector $\bq$ is continuous and independent on $V$ (in contrast
  to the finite torus), and ($\lim_{\bq\to 0}, \lim_{V\to \infty})$ commute.
  The complex frequency $\omega+i\ve\to 0$, can be taken (after $V\to \infty$)
  along the imaginary axis. }.

\bibitem{Comm-OBC}
{On OBC one can define {\em uniform} charge and thermal polarizations, and the
  charge and thermal magnetizations needed for Eq.~(\ref{Kubo}). Continuous and
  uniform thermal gradients can only be implemented on OBC. For the mesoscopic
  regime, boundary conditions matter. Our formulas can be relevant for small
  mesoscopic samples where the dephasing length scales is of the order of the
  sample size. Our finite volume formulas can implement the effects of attached
  leads, by keeping $\varepsilon$ larger than the leads level spacing.}

\bibitem{Comm-C4}
{ For models with no C4 symmetry, the Hall coefficients, which are
  antisymmetric in magnetic field $B$, can be defined by antisymmetrizing the
  Kubo formulas with respect to $j_\so^x\to j^y_{\so'}$, in accordance with
  Onsager's relations.}

\bibitem{Comm-SD}
{For long range interactions, the spatial decomposition may not be unique.}

\bibitem{Comm-Particles}
{In the literature~\cite{Cooper}, one frequently encounters specialized
  expressions of the magnetization and thermal magnetization $M_{\so\so'}$ for
  continuum particle Hamiltonians, $M^{\rm p}_{\sc\st} = -{e\over 2 }
  \sum_i^{N_p} \hat{\bz} \cdot \bx_i \times \bv_i $, and $M^{\rm p}_{\st\st}=
  -{1\over 2 } \sum_i^{N_p} \hat{\bz} \cdot \bx_i \times \{ \bv_i , h_i\},$
  where $\bv=(\bp_i -{e\over c} \bA_i)/m$. Here we consider more general
  lattice Hamiltonians.}

\bibitem{Read-Kubo}
Barry Bradlyn and N~Read.
\newblock Low-energy effective theory in the bulk for transport in a
  topological phase.
\newblock {\em Physical Review B}, 91(12):125303, 2015.

\bibitem{Comm-anomalous}
{We describe normal diagonalized hamiltonians. Anomalous bosonic terms e.g.
  $a^\dagger a^\dagger$ may be present in the energy-diagonalized thermal
  polarizations. These terms contribute off-shell matrix elements which can be
  ignored.}

\bibitem{Comm-smooth}
{The following argument holds also for sharp and well separated edges in the
  bulk-gapped QH phases. }.

\bibitem{Ziman}
John~M Ziman.
\newblock {\em {Electrons and phonons: the theory of transport phenomena in
  solids}}.
\newblock Oxford university press, 2001.

\bibitem{Lauchli}
Andreas~M. L\"auchli, Julien Sudan, and Erik~S. S\o{}rensen.
\newblock {Ground-state energy and spin gap of spin-$\frac{1}{2}$
  Kagom\'e-Heisenberg antiferromagnetic clusters: Large-scale exact
  diagonalization results}.
\newblock {\em Phys. Rev. B}, 83:212401, Jun 2011.

\bibitem{DMRG}
Ulrich Schollwock.
\newblock {The density-matrix renormalization group in the age of matrix
  product states}.
\newblock {\em Annals of Physics}, 326(1):96 -- 192, 2011.
\newblock January 2011 Special Issue.

\bibitem{SM}
{See Supplementary Material}.

\bibitem{Abhisek}
Abhisek Samanta, Daniel~P. Arovas, and Assa Auerbach.
\newblock {Hall Coefficient of Semimetals}.
\newblock {\em Phys. Rev. Lett.}, 126:076603, Feb 2021.

\bibitem{CC}
JT~Chalker and PD~Coddington.
\newblock Percolation, quantum tunnelling and the integer hall effect.
\newblock {\em Journal of Physics C: Solid State Physics}, 21(14):2665, 1988.

\end{thebibliography}

\end{document}